\documentclass[twocolumn,aps,prl,10pt,amsmath,amssymb,nofootinbib,showpacs,superscriptaddress,floatfix]{revtex4-1}

\DeclareFontFamily{U}{rcjhbltx}{}
\DeclareFontShape{U}{rcjhbltx}{m}{n}{<->rcjhbltx}{}
\DeclareSymbolFont{hebrewletters}{U}{rcjhbltx}{m}{n}

\usepackage{graphicx}
\usepackage{color}
\usepackage[usenames,dvipsnames]{xcolor}
\usepackage[colorlinks=true,linkcolor=Red,citecolor=Green,linktoc=page]{hyperref}
\usepackage{multirow}
\usepackage{float}
\usepackage{flushend}
\usepackage{balance}
\usepackage[varg]{txfonts}
\usepackage{ulem}
\usepackage{fancyhdr}

\DeclareMathSymbol{\lamed}{\mathord}{hebrewletters}{108}

\begin{document}
\title{Room-Temperature Superconductivity in 1D}




\author{C.\,A.\,Trugenberger}

\affiliation{SwissScientific Technologies SA, rue du Rhone 59, CH-1204 Geneva, Switzerland}


\
\begin{abstract}
We review the theoretical model underpinning the recently reported room-temperature, ambient-pressure superconductivity along line defects on the surface of highly-oriented pyrolytic graphite. The main ingredients for this 1D room-temperature superconductivity are pairing by effective strain gauge fields, the formation of an effective Josephson junction array in its Bose metal state on the surface and the suppression of phase slips by dimensional embedding in an extremely well-conducting 3D bulk structure.
\end{abstract}
\maketitle

\section{Introduction} 
\begin{figure}[H]
\includegraphics[width=5.5 cm]{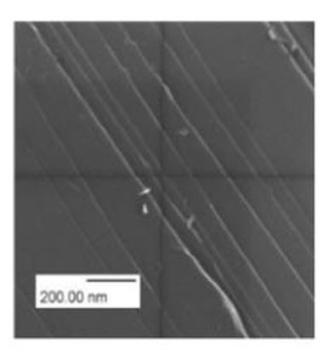}
\caption{The
 cleaved surface of the HOPG sample, clearly showing the nearly parallel defect lines. \label{fig1}}
\end{figure}

Recently, we have reported ambient-pressure, room-temperature global superconductivity in scotch-tape-cleaved, highly oriented pyrolytic graphite (HOPG) \cite{ref-kopelevich}. Cleaving thin slabs of HOPG produced many nearly parallel defect lines on the surface, as shown in Figure~\ref{fig1}. Superconducting currents flow through these 1D defects. Superconductivity was established by the simultaneity of resistance drops and the onset of magnetic screening. 
Its global character was measured by matching local and non-local resistance curves. 

\vspace{-9pt}

This superconductivity along line defects on the HOPG surface seems to defy the conventional wisdom that superconductivity cannot exist in 1D (for a review, see \cite{ref-arutyunov}). In this contribution, I will focus exclusively on the theoretical aspects of the mechanism which does, indeed, allow superconductivity in 1D, albeit only in a very particular configuration. In a nutshell, superconductivity in 1D can be realized if the lines are edges of a granular bosonic topological insulator \cite{ref-lu}, so that Cooper pairs are symmetry-protected from dissipation and by suppressing quantum phase slips so as to achieve global phase coherence on the 1D line defects. In the following, we will come back in detail to these points. But first, we explain how the HOPG electrons pair up along the 1D defects. 

\section{Pairing by Strain Gauge Fields} 
Defects represent strain lines on the surface of the HOPG slab. The abundant electrons on the HOPG layers feel strain as an effective 2D gauge field (for a review, see \cite{ref-guinea}). The gauge-invariant fields that really influence electron behavior are actually strain gradients, and in particular, the defects act as external magnetic fields that localize electrons in granules along them, where the localization along the defects is due to shear gradient strain and the formation of droplets along the defects is due to modulations of diagonal strain \cite{ref-kopelevich}. Since electrons are spin-polarized by the effective magnetic field, this gives rise to the typical local ferromagnetism along defects observed on HOPG surfaces \cite{ref-cervenka}. 

There is another possible state competing with ferromagnetism, though. It is known that random strain fluctuations constitute the major source of disorder on HOPG sheets and that the in-plane fluctuations are the dominant ones \cite{ref-couto}. These are represented by fluctuating gauge fields coupling to the electrons \cite{ref-couto, ref-kopelevich}. The presence of line defects on the surface, in this context, has a crucial effect. Since the line defects explicitly break the 2D parity symmetry on the surface, the effective gauge action for the strain gauge field $a_{\mu}$ must contain, as the infrared-dominant term, the Chern--Simons term \cite{ref-jackiw},
\begin{equation} 
S_{\rm strain}= \int d^3 x \ {\kappa\over 4\pi} a_{\mu} \epsilon^{\mu \alpha \nu} \partial_{\alpha} a_{\nu} -{1\over 4 g^2} f_{\mu \nu}f^{\mu \nu} \ ,
\label{csaction}
\end{equation}
where the Greek letters denote the components of 3D Minkowski space with coordinates $x=(vt, {\bf x})$, with $v=O\left(10^{-2}\right) c$ being the velocity of light in graphene, Einstein notation is used and $\epsilon^{\mu \alpha \nu}$ is the totally antisymmetric tensor. The quantity $f_{\mu \nu}=\partial_{\mu} a_{\nu}-\partial_{\nu}a_{\mu}$ is the effective field strength tensor, $\kappa$ is a dimensional effective coupling and $g^2$ is an effective coupling with the canonical dimensions mass (we use natural units $c=1$, $\hbar = 1$, $\varepsilon_0=1$). The Chern--Simons term (first term in the action) gives a gauge-invariant mass $\mu = |\kappa| g^2/2\pi$ to the gauge field. This means that, in the presence of the defects, strain fluctuations are screened by this mass and become short-range. 

Due to the Chern--Simons term, an electron at rest is the source of not only an effective electric field but also an effective magnetic field. The Pauli interaction of this field with the spin of the second electron results in an attractive component of the pair potential. The two-body potential mediated by gauge fields, including the Chern--Simons term for aligned spin 1/2 electrons of mass $m$, has been derived in \cite{ref-kogan, ref-semenoff},
\begin{equation} 
V = {\left( L -{1\over \kappa} + {\mu v r \over \kappa} K_1\left( \mu v r \right) \right)^2 \over mr^2 } -{\mu v^2 \over \kappa} {\mu-m \over m} K_0 \left( \mu v r \right) 
+ V_{\rm C} (r) \ ,
\label{2body}
\end{equation}
where we have added the electric Coulomb potential $V_{\rm C} = (qe)^2/(4\pi \epsilon r)$ and $K_{0,1}$ denote modified Bessel functions of the second kind, with the short distance asymptotic behaviors $K_0(x) \propto -{\rm ln} x$, $K_1(x) \propto 1/x$, while both functions are asymptotically suppressed at large distances.
Due to Fermi statistics, the integer $L$, representing twice the angular momentum, can take only odd values $L=2k +1$, $k \in {\mathbb Z}$. 

At large distances, $r \gg 1/v\mu$, only the Chern--Simons term survives. Here, the pair potential reduces to the electric Coulomb potential plus a centrifugal barrier with a modified angular momentum $(1/2)(L-1/\kappa )$. For $\kappa=1/{\rm even \ integer}$, the angular momentum spectrum is unchanged; otherwise, the Chern--Simons term induces a statistical transmutation, which is reflected in an anomalous magnetic moment. In terms of spin 1/2 electrons, this can be interpreted as an anomalous gyromagnetic factor. These observations are consistent with previous measurements of an intrinsic magnetic field and an anomalous gyromagnetic factor when the parity symmetry is broken by an external magnetic field \cite{ref-sercheli, ref-schneider}. 

At short distances, however, the Maxwell term also plays a role. Specifically, for small $\kappa$ and $\mu > m$, the magnetic attraction due to the anomalous magnetic moment always dominates both the effective and the real Coulomb repulsion, and there exist bound states \cite{ref-kogan}. The smaller $\kappa$, the larger the angular momentum range for which the centrifugal barrier is canceled or becomes attractive. Therefore, for small enough $\kappa$, we can have spin-triplet bound states in the $s-$, $p-$ and even $d-$wave states. It must be considered, however, that the coupling $g^2$ is renormalized by the interaction with the electrons. In the relativistic theory, at one loop level, this is given by \cite{ref-semenoff},
\begin{equation}
{1\over g^2} = {1\over g_0^2} + {1\over 12 \pi m} \ ,
\label{renor}
\end{equation}
where $g_0^2$ is the bare value. This implies that, for large bare coupling $g_0^2$, the renormalized gauge field mass becomes $\mu = 6\kappa m$. The condition
$\mu > m$, therefore, reduces to $\kappa > 1/6$. This excludes pairings with angular momentum higher than the d-wave. 

Using the above asymptotic expansions of the modified Bessel functions of the second kind, it is easy to see that the minimum of the pair potential is at distances 
$O(1/v\mu)$ or smaller. In our case, the length scale of strain fluctuations is given by the line defect width $w$, which implies $\mu = O(1/vw)$. Using the typical defect width $w = 10 {\rm nm}$ and the light velocity in graphite $v=10^6 {\rm m/s}$, we obtain an energy scale corresponding to a transition temperature $T_{\rm C} = O(1000) ^{\circ}K$. It is the small lateral dimension of the defects that is responsible for the elevated pairing temperature. 

\section{Josephson Junction Arrays and Bose Metals} 

\begin{figure}[H]
\includegraphics[width=7.5 cm]{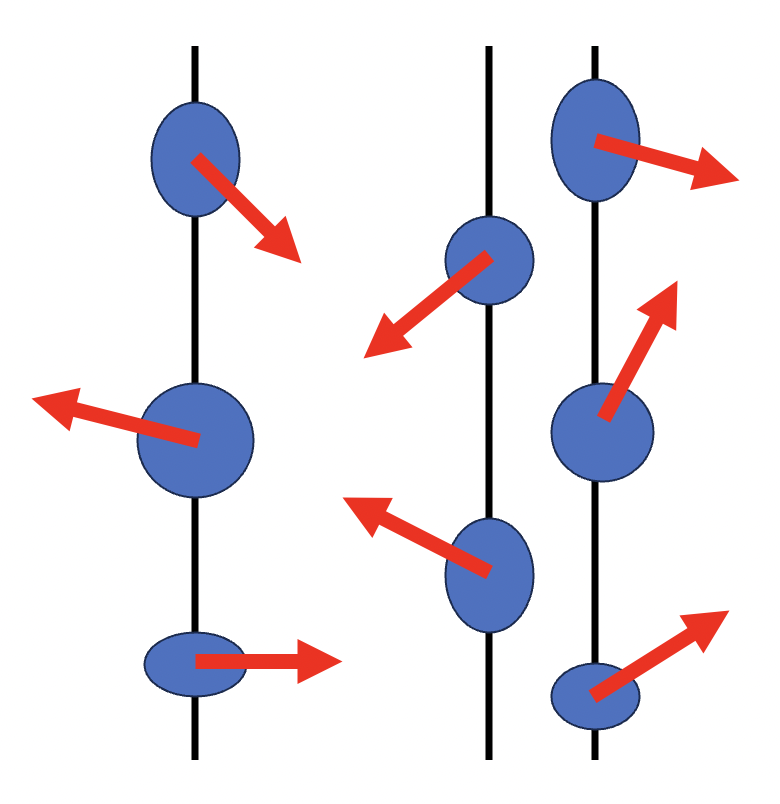}
\caption{A schematic representation of the effective JJA forming on the cleaved surface of the HOPG sample. The black lines denote the defects, the blue dots represent the droplets of condensate along them and the red arrows represent the phases of local condensates. \label{fig2}}
\end{figure}

As a consequence of the strain gauge field interaction, electrons localized on droplets along the 1D defects can pair and condense there. The result is the formation of an emergent Josephson junction array (JJA) (for a review, see, e.g., \cite{ref-zant}) on the surface of the HOPG sample, as shown schematically in Figure~\ref{fig2}. 

\begin{figure}[H]
\includegraphics[width=8.5 cm]{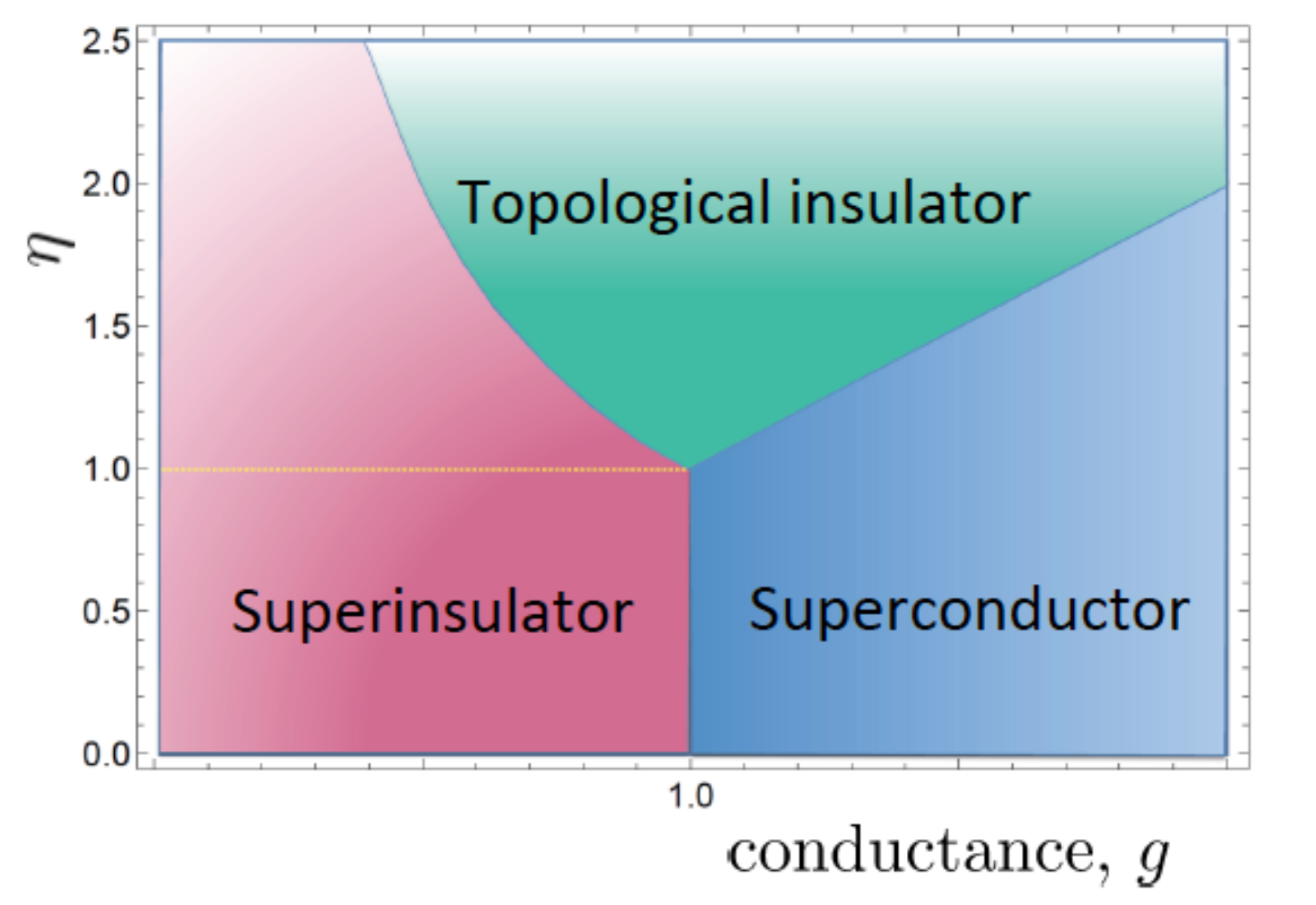}
\caption{The quantum phase structure of Josephson junction arrays. The two parameters $g$ and $\eta$ are the two dimensionless quantities that can be constructed from the array magnetic and electric couplings and from the distance between superconducting droplets. \label{fig3}}
\end{figure}

The correct phase structure of a JJA and thin superconducting films, which are best modeled as an emergent JJA, as in this case, was first derived in \cite{ref-dst} (for a review, see \cite{ref-book}) and is driven by two quantum parameters at low temperatures. To see this, let us recall that, for a JJA, the electric coupling is a dimensionful quantity of canonical dimension [mass] given by $E_C = O\left( e^2/C \right)$, where $e$ is the electron charge and $C$ represents the droplet capacitance. Since the Josephson coupling $E_{\rm J}$ between the droplets is also dimensionful, with canonical dimension [mass], we can construct one dimensionless parameter $g$ as the ratio of these magnetic and electric couplings and one parameter with canonical dimension [mass] as the square root of their product. We then have a second dimensionless parameter $\eta$, which is the product of this mass parameter and the typical inter-droplet distance. Broadly speaking, $g$ drives the quantum transition from a superconductor at high values to a superinsulator (for a review, see \cite{ref-book}) at low values. However, if $\eta >1$, an intermediate phase can open up between the two, as shown in Figure~\ref{fig3}.

This new phase of matter, nowadays often called a Bose (or anomalous) metal, was first predicted in \cite{ref-dst}. Although the charges are Cooper pairs, it is characterized by a metallic saturation of the resistance at low temperatures, hence the name “Bose metal”. The zero-temperature sheet resistance is symmetric around the resistance quantum $R_{\rm Q} = h/4e^2 = 6.45 \ k\Omega$, with $R_\square < R_{\rm Q}$ for $g>1$ and $R_\square > R_{\rm Q}$ for $g<1$. The former region is often also called a ``failed superconductor", while the latter domain is dubbed a ``failed insulator”. 

The Bose metal is \cite{ref-bm} actually a bosonic topological insulator \cite{ref-lu, ref-senthil}, in which Cooper pairs and vortices are frozen by their statistical interactions in the bulk but can move freely on edges, forming a percolation network through the sample \cite{ref-cc}. Cooper pairs move along the edges, while vortices traverse them perpendicularly, thereby generating the dissipation mechanism leading to the metallic behavior via quantum phase slips \cite{ref-arutyunov}. 
The Bose metal intermediate phase is now clearly seen in both thin superconducting films, as seen in, e.g., \cite{ref-kapitulnik}, and JJAs \cite{ref-marcus1}. Moreover, a recent experiment \cite{ref-marcus2} has confirmed that, in this phase, vortices are out-of-condensate and frozen, apart for a small fraction that can move freely, which is perfectly consistent with the bosonic topological insulator picture originally advocated in \cite{ref-dst, ref-bm}. 

\begin{figure}[H]
\includegraphics[width=8.5 cm]{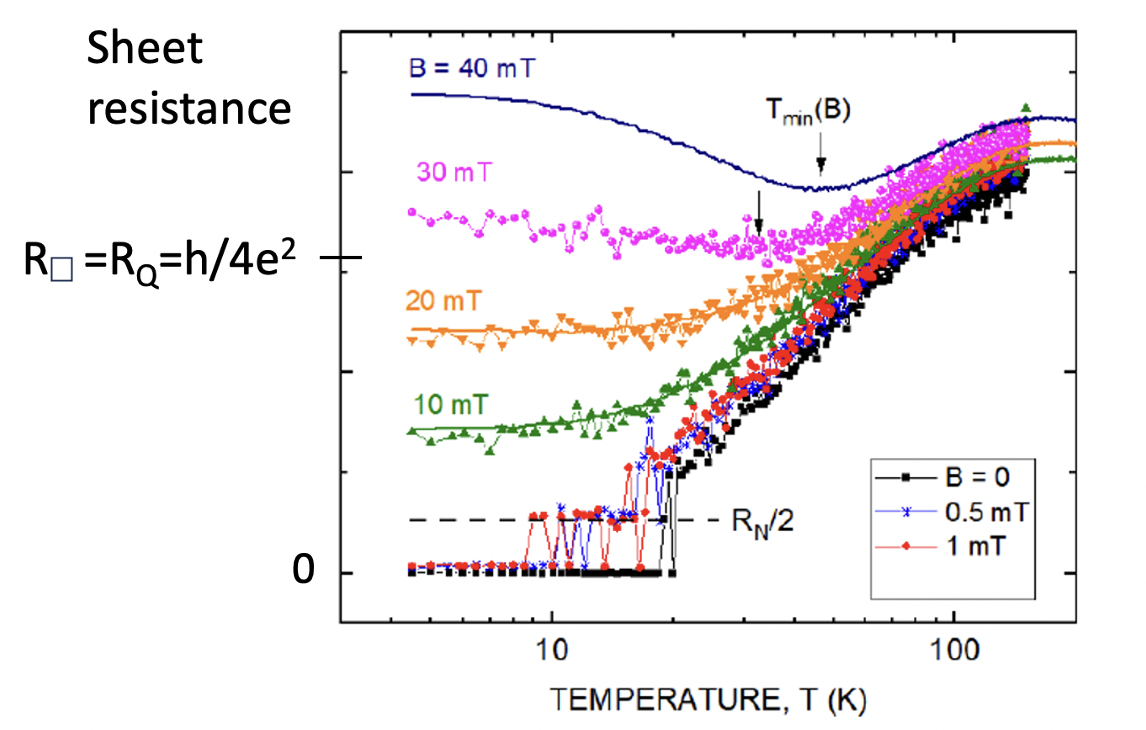}
\caption{The Bose metal state of the emergent JJA on the cleaved surface of the HOPG sample. Note the metallic saturation of the resistance, symmetric around the resistance quantum. The transition between a failed superconductor and the failed insulator is driven here by an external magnetic field. At the lowest values of $B$, the failed superconductor does not fail. \label{fig4}}
\end{figure}

The emergent JJA on the HOPG surface is in its Bose metal state, the line defects assuming the role of edges in the Chalker--Coddington picture described above \cite{ref-cc}, as shown in Figure~\ref{fig4}. The quantum phase transition from failed superconductor to failed insulator is driven here by an external magnetic field, which takes the role of the parameter $g$ described before.

The crucial point, however, is that, at the lowest magnetic fields, the failed superconductor has not failed. At a given $T(B)$, there is an abrupt drop in the resistance to zero. It looks like quantum phase slips, the only possible dissipation mechanism on the edges of a bosonic topological insulator, are suppressed in this region. To understand how this suppression comes about, let us revisit the theory of Josephson junction chains (JJCs) on  \mbox{the edges.} 

\section{Josephson Junction Chains and Quantum Phase Slips}

Josephson junction chains are 1D arrays of spacing $\ell$ of superconducting islands with nearest neighbors Josephson couplings of strength $E_J$. Each island has a capacitance $C_0$ to the ground and a mutual capacitance $C$ to its neighbors. The Hamiltonian for such a system is
\begin{equation}
H=\sum_{\bf x} \ {C_0\over 2} V_{\bf x}^2 + \sum _{<{\bf x \bf y}>}
\left( {C\over 2} \left( V_{\bf y}-V_{\bf x} \right) ^2 + E_J
\left( 1-{\rm cos}\ \left( \varphi _{\bf y} - \varphi _{\bf x} \right) \right)
\right) \ ,
\label{ham1}
\end{equation}
where boldface characters denote the sites of the array, $<{\bf x \bf y}>$ indicates nearest neighbors, $V_{\bf x}$ is the electric potential of the island at ${\bf x}$ and $\varphi _{\bf x}$ is the phase of its order parameter.  

Using the notation $\Delta$ and $\hat \Delta$ for forward and backward finite differences on the 1D array, respectively, this Hamiltonian can be rewritten as
\begin{equation}
H= \sum_{\bf x} \ {1\over 2} V \left( C_0 - C\nabla^2 \right) V +
\sum _{{\bf x},i} \ E_J \left( 1-{\rm cos} \  \left( \Delta \varphi  \right)
\right) \ ,
\label{ham2}
\end{equation}
where $\nabla^2 = \hat \Delta\Delta $ is the discrete Laplacian, and we have omitted, for simplicity of presentation,  the location indices on the potential and phase variables. 

The phases $\varphi $ are quantum-mechanically conjugated to the charges (Cooper pairs) $2e q$, $q \in {\mathbb Z}$, on the islands, where $e$ is the electron charge. The Hamiltonian (\ref{ham2}) can be expressed in terms of charges and phases by noting that the electric potentials $V$ are determined by the charges $2e q$ via a discrete version of Poisson's equation:
\begin{equation}
\left( C_0 -C\nabla^2 \right) V= 2e q \ .
\label{poisson}
\end{equation}

Using this in (\ref{ham2}), we obtain
\begin{equation}
H= \sum_{\bf x} \ {2 e^2\over C}\ q{1\over {C_0/C}-\nabla^2 } q + E_{\rm J} \left( 1-{\rm cos} \ \left( \Delta  \varphi \right)
\right) \ .
\label{hamfin}
\end{equation} 
The interaction between charges is, thus, screened on the scale $\sqrt{C/C_0} $. Since the JJC lives on the edge of a JJA in its topological insulator state, in which bulk charges and vortices are frozen and Coulomb interactions are very short range, we will consider the limit $C<C_0$ in which the screening length of the edge Coulomb interaction is less than a lattice spacing, and we will, consequently, treat electric interactions as diagonal. In this case, the Hamiltonian reduces to
\begin{equation}
H= \sum_{\bf x} \ 4E_{\rm C} \ q^2 + E_{\rm J} \left( 1-{\rm cos} \ \left( \Delta  \varphi \right) \right)\ ,
\label{add}
\end{equation}
where $E_{\rm C} = e^2/2C_0 $. This charging energy and the Josephson coupling $E_{\rm J} $ are the two relevant energy scales in the problem. These can be traded for the Josephson plasma energy $E_{\rm P} = \sqrt{8E_{\rm C} E_{\rm J}}$ and the dimensionless coupling $g = \sqrt{E_{\rm J} / 32 E_{\rm C}}$ encoding the ratio of magnetic to electric couplings. Note that these are not the same parameters as those of the surface 
JJA since, there, long-range 2D Coulomb interactions dominate, while here, on the edges, 1D Coulomb interactions are very short-range when the surface is in its Bose metal state. 

The zero-temperature partition function of the Josephson junction chain admits a path-integral representation, which can be easily written down in first-order form \cite{ref-fazio, ref-dst}. Since the charge variables $q$ are integers, time integration has to be performed stepwise. We, therefore, introduce a lattice spacing $\ell_0$ also in the time direction. This represents the typical time scale on which charges change by an integer step. The Euclidean time partition function is given by
\begin{eqnarray}
Z &&= \sum_{\{q \}} \int_{-\pi }^{+\pi } {\cal D}\varphi \ {\rm exp}(-S)\ ,
\nonumber \\
S &&=\sum_{x}  -i\ q \Delta_0 \varphi + 4 E_{\rm C} \ell_0 \ q^2 + \ell_0 E_{\rm J}\ \left( 1- {\rm cos}\ \left( \Delta \varphi \right)
\right)\ ,
\label{partfunc}
\end{eqnarray} 
where $\Delta_0$ is the finite difference operator in the time direction and the sum in the action $S$ extends over the 2D lattice with spacing $\ell_0$ in the Euclidean time direction. In the next step, we replace the Josephson term with its Villain form (for a review, see \cite{ref-kleinert}), 
\begin{eqnarray}
Z &&=\sum_{\{q, v \}} \int_{-\pi }^{+\pi } {\cal D}\varphi
\ {\rm exp}(-S)\ ,
\nonumber  \\
S&&=\sum _{x}  -i\ q \Delta_0 \varphi + 4 \ell_0 E_{\rm C} \ q^2
+ \ell_0 {E_{\rm J} \over 2} \left( \Delta \varphi +2\pi v_i \right)^2 \ ,
\label{part}
\end{eqnarray} 
with new integer variables $v$ defined on the spatial links of the 2D array. Finally, we use the Poisson formula
\begin{equation}
\sum_{k=-\infty}^{k=+\infty} {\rm exp} (i2\pi k z) = \sum_{n=-\infty}^{n=+\infty} \delta (z-n) \ ,
\label{poisson}
\end{equation}
to transform $q$ into a real variable at the price of introducing a new set of integer dynamical variables $v_0$ defined on the Euclidean time links. By grouping together finite differences and v variables into two vectors $\Delta_\mu$ and $v_{\mu}$, where $\mu = 0, 1$, we can write the partition function as
\begin{eqnarray}
Z &&=\sum_{\{v_\mu\}} \int_{-\infty}^{+\infty} {\cal D} q \ \int_{-\pi }^{+\pi } {\cal D}\varphi
\ {\rm exp}(-S)\ ,
\nonumber  \\
S&&=\sum _{x}  -i\ q \left( \Delta_0 \varphi +2\pi v_0 \right)  + 4 \ell_0 E_{\rm C} \ q^2
+ \ell_0 {E_{\rm J} \over 2} \left( \Delta_1  \varphi +2\pi v_1 \right)^2 \ .
\label{part}
\end{eqnarray} 

Finally, we integrate explicitly over $q$, to obtain
\begin{eqnarray}
Z &&=\sum_{\{v_\mu\}} \int_{-\pi }^{+\pi } {\cal D}\varphi
\ {\rm exp}(-S)\ ,
\nonumber  \\
S&&=\sum_{x} {1\over 16 \ell_0 E_{\rm C}} \left( \Delta_0 \varphi +2\pi v_0 \right)^2 
+ \ell_0 {E_{\rm J} \over 2} \left( \Delta _1 \varphi +2\pi v_1\right)^2 \ .
\label{part}
\end{eqnarray} 

Using the previously defined couplings, the action in this partition function can be written in the form
\begin{equation}
S= g \sum_{x}  {1\over \ell_0 E_{\rm P}} \left( \Delta_0 \varphi+2\pi v_0 \right)^2 
+ \ell_0 E_{\rm P} \left( \Delta _1 \varphi +2\pi v_1\right)^2 \ ,
\label{finac}
\end{equation}
which defines a non-relativistic version of the compact O(2) model in (1+1) dimensions \cite{ref-polyakovbook}. 

To proceed, we note that the combination $\ell_0 E_{\rm P}$ plays the same role on the edges as the previously defined parameter $\eta$, which co-determines the quantum phase structure of the emergent JJA on the HOPG surface. The plasma frequency $E_{\rm P}$ is the ultraviolet cutoff in the present model; at higher energies, plasma oscillations can be excited. We, thus, consider time scales satisfying $\ell_0 E_{\rm P}  \gtrsim 1$, for which the model becomes essentially relativistic. We then follow \cite{ref-polyakovbook}, and we decompose the compact O(2) action as
\begin{equation}
S= g \sum_{x, \mu} \Delta_\mu \varphi \Delta_\mu \varphi + 4\pi^2 g \sum_x m {1\over -\nabla^2} m  \ ,
\label{decomp}
\end{equation}
where the integers $m =\epsilon_{\mu \nu} \Delta_{\mu} v_{\nu}$ represent the quantum phase slip instantons \cite{ref-polyakovbook}. The first term in this decomposition encodes the symmetry-protected charged boson propagating on the edge, with electric current $j_{\mu} \propto 2e \Delta_\mu \varphi$.  In the absence of instantons, this current is conserved, which means there is no dissipation. The only possible source of dissipation lies in the quantum phase slip instantons, which can disorder the system. Importantly, however, quantum phase slip instantons are not independent of each other; they form a 2D Coulomb gas, like vortices in the XY model (for a review, see \cite{ref-minnhagen}). Therefore, they undergo the Berezinskii--Kosterlitz--Thouless transition (for a review, see \cite{ref-minnhagen}), not as a function of temperature but rather of quantum coupling $g$. At sufficiently large values of $g$, quantum phase slip instantons are suppressed since being bound in neutral pairs, and the edge becomes superconducting. It remains to be seen how we can achieve such a significantly large coupling. 

\section{Suppressing Quantum Phase Slips by Embedding in Higher Dimensions}

\begin{figure}[H]
\includegraphics[width=8.5 cm]{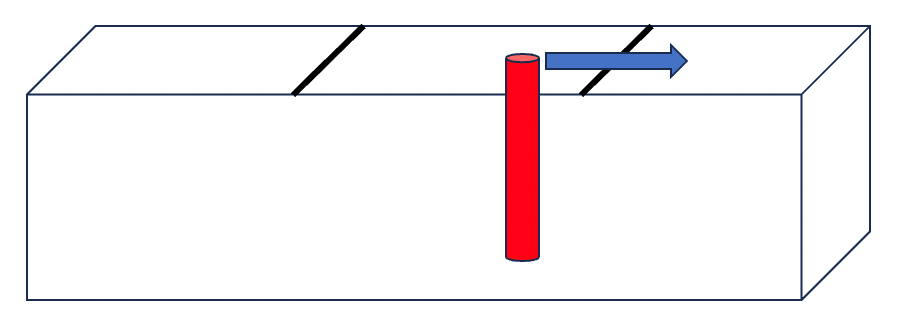}
\caption{A bulk vortex moving, thereby generating a quantum phase slip on the line defect on the surface. \label{fig5}}
\end{figure}

The key observation for the suppression mechanism is that, in our case, the JJC is not a standalone entity, but rather, it is embedded in a higher-dimensional structure. First of all, the JJC is the edge of a JJA on a 2D surface. As a consequence, a quantum phase slip, a tunneling event corresponding to a flip by $2\pi$ of the phase on one droplet of the JJC on the time scale $\ell_0$, when viewed from the 2D point of view, corresponds to a vortex traversing the JJC between two droplets. But the surface on which the emergent JJA lives is, in turn, the boundary of a 3D sample. Therefore, the traversing 2D vortex is actually the endpoint of a vortex line in the bulk sample, as shown in Figure~\ref{fig5}.

The bulk of HOPG is an extremely good conductor---its resistivity is $0.4 * 10^{-8} \ \Omega m$, which is smaller than the resistivity $1.59 * 10^{-8} \ \Omega m$ of even silver, the best metal conductor. In such an extremely good conductor, the motion of vortices is severely suppressed, with the consequence that also quantum phase slips in the line defects on its surface are quashed and these defects 
become superconducting. Moreover, since this bulk resistivity remains so small up to room temperature, thermal phase slips are also absent up to very high temperatures. 

\section{Conclusion: The Ingredients}
There are two ingredients of 1D room-temperature superconductivity. First of all, one needs very thin strain line defects on the surface of a bulk sample so that strain fluctuations on this surface can lead to pairing in droplets along the defects. Secondly, the bulk material must be an excellent conductor, so as to suppress vortex motion and, by dimensional cascade, phase slips along the emergent JJC in the line defects.

\end{document}